# The horizontal profile of the atmospheric electric fields as measured during thunderstorms by the network of NaI spectrometers located on the slopes of Mt. Aragats

*A. Chilingarian, G. Hovsepyan, T.Karapetyan, L. Kozliner, S.Chilingaryan. D.Pokhsraryan, and B.Sargsyan.*

Yerevan Physics Institute, Alikhanyan Brothers 2, Yerevan 0036, Armenia

**Abstract**

The shape and evolution of the energy spectra of the thunderstorm ground enhancement (TGE) electrons and gamma rays shed light on the origin of TGEs, on the relationship between modification of the cosmic ray electron energy spectra (MOS) and relativistic runaway electron avalanche (RREA) processes, on the energy of the seed electrons, and on the strength and elongation of an atmospheric electric field. The network of large NaI spectrometers on slopes of Mt. Aragats 24/7 monitored secondary particle fluxes from 2013 until now, highly contributed to the understanding of the ways how RREAs are developed in the atmosphere. In 2022 we enlarge the NaI network with 2 remote detectors located at altitudes 2000 and 1700 m, and 13 and 16 km apart from the Aragats station to investigate the horizontal profile of the atmospheric electric field. We found, that the previously estimated values of the regions in the atmosphere, where RREA emerges, were highly underestimated. In the present report, we describe the NaI particle detector's network and present the first results of the experiment demonstrating that the particle fluxes from the atmospheric electron accelerators can cover large areas on the earth's (up to tens of km$^2$).

1. **Introduction**

Gamma-ray instrumentation for the high-energy physics in the atmosphere (HEPA) consists of networks of large area scintillation and NaI (TL) spectrometers combined with sophisticated post-processing of registered energy releases. The main difficulty of the experiments for gamma-ray spectroscopy on the earth's surface is distinguishing between gamma-ray emission of Radon progeny, intensified by disturbances of the near-surface electric field, and gamma rays from the electron-gamma ray avalanches unleashed in strong atmospheric electric fields. The only possibility for the reliable detection of gamma rays from the electron accelerators operated in the thunderous atmosphere is a significant enlargement of the spectrometer range. Most of the spectrometers available from the industry have a very small range of 0.3-3 MeV. Thus, we fabricate spectrometers with a range of 0.3 – 100 MeV for registration of thunderstorm ground enhancements (TGEs) of the highest energies and reliably distinguishing them from the Radon progeny gamma radiation.

The network of gamma spectrometers operated on slopes of Mt. Aragats allows for solving the most imporatntproblems of HEPA, including vertical and horizontal profiles of the atmospheric electric field during thunderstorms. The size of the particle emitting region in a thundercloud still remains not well researched. Measurements with multiple dosimeters installed at nuclear power plants in a coastal area of the Japanese sea made it possible to follow the source of the gamma ray flux moving with an ambient wind flow [1]. At Nor Amberd research station, located on slopes of the Mt. Aragats at 2000 m height, the size of the particle emitting region was estimated Using the muon stopping effect [2,3]. Estimates from both studies locate particle emitting regions within 1 km. However, in the recent radar-based gamma glow (TGE) study along the coast of the Japanese sea, it was observed that all TGEs were accompanied by the graupel fall, indicating the low location of the lower positively charged region [4]. A strong radar echo due to the high reflectivity of hydrometeors indicates that the vertical and horizontal extent of the strong accelerating electric field was larger than 2 km. In another observation of the gamma glow in Japan the flux enhancements were initiated and terminated exactly at the same time at a distance of 1.35 km [5]. Thus, the previously estimated values of particle emitting region size within 1 km seem to be highly underestimated.

## 2. Spectrometry of gamma radiation on Mt. Aragats

The NaI network is located in on slopes of Mt. Aragats at heights from 1700 to 3200 m, see Fig 1. It consists of 8 homemade spectrometers based on large (12 x 12 x 24 cm) NaI(TL) crystal scintillators [6], and 7.6×7.6 cm diameter and length, 1024 channels ORTEC spectrometer with excellent energy resolution $\approx 7\%$, however with very limited energy range from 0.3 to 3 MeV [7]. Although the energy resolution of home-made spectrometers is rather low 50-60% at 662 KeV comparing with Orteg spectrometer, the big advantage of spectrometers is the possibility to measure low intensity flux of the highest energies of TGE particles (up to 50 MeV). Orteg spectrometer was used for the Radon progeny gamma radiation spectroscopy in the energy range of 0.3 – 3 MeV. The energy deposits collection time is one minute, count rates are available also on 1s time span. From the 1-minute time series of the recovered energy spectra we can see the dynamics of the developing TGE and its contraction after a lightning flash. 5 large NaI spectrometers are located under the roof in the SKL experimental hall at Aragats high altitude research station (3200 m), 2 of them have energy threshold 0.3 MeV, other 2 - 3 MeV for the better measurements of highest energies. One of spectrometers is attached to high-speed digitizing oscilloscope for comparing pulses from atmospheric discharges and particle pulses on a nanosecond time scale. 2 spectrometers are located in Nor Amberd research station and one – in Burakan (see Fig.1).

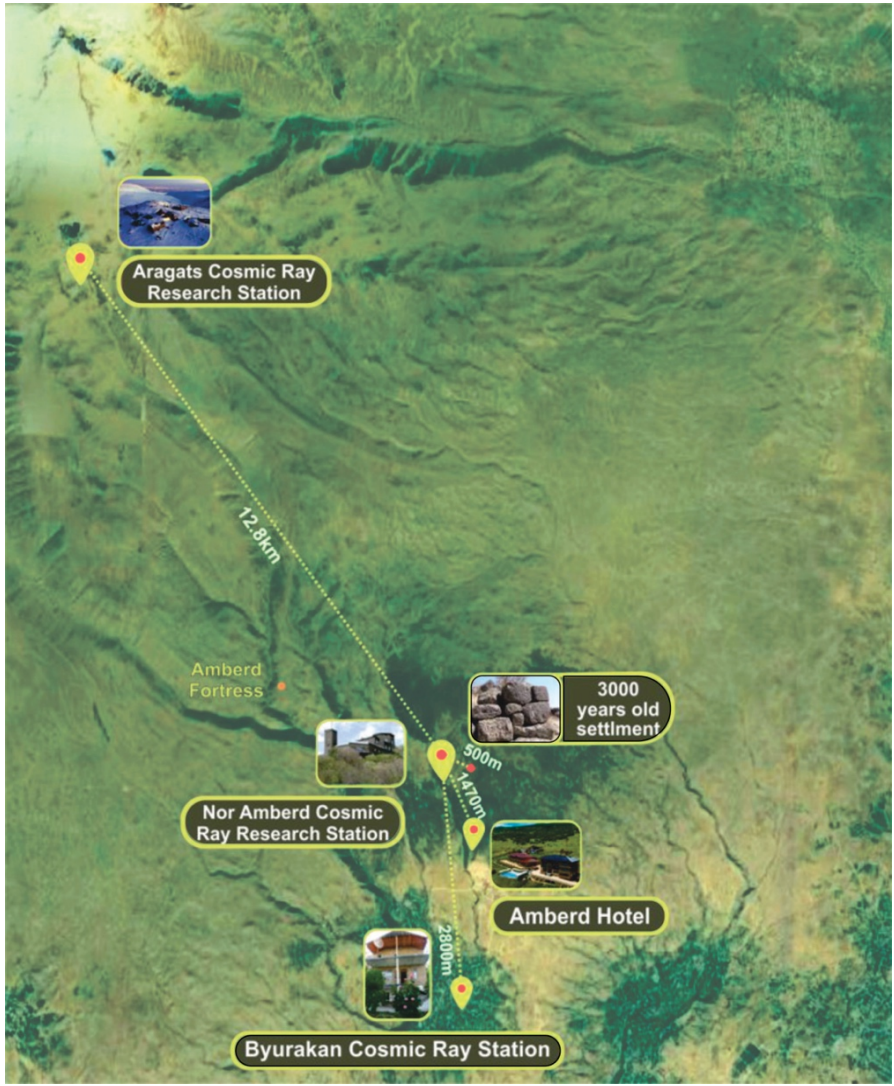

**Figure 1. The map of networks of NaI (spectrometer locations: five on Aragats (3200 m), one in Burakan (1700 m), and one in Nor Amberd station (2000 m). Electric mills and lightning locators are installed on Aragats (5 units) and in Nor amberd.**

The NaI crystal is surrounded by 0.5 cm of magnesium on all sides (because the crystal is hygroscopic) with a transparent window directed to the photocathode of the photomultiplier tube PM-49, see Fig. 2. The large cathode of PM-49 (15-cm diameter) provides good light collection. The spectral sensitivity range of PM-49 is 300–850 nm, which covers the spectrum of NaI(Tl) emission light. The sensitive area of each NaI crystal is ~0.032 m$^2$; the total area of the five crystals is ~0.16 m$^2$.

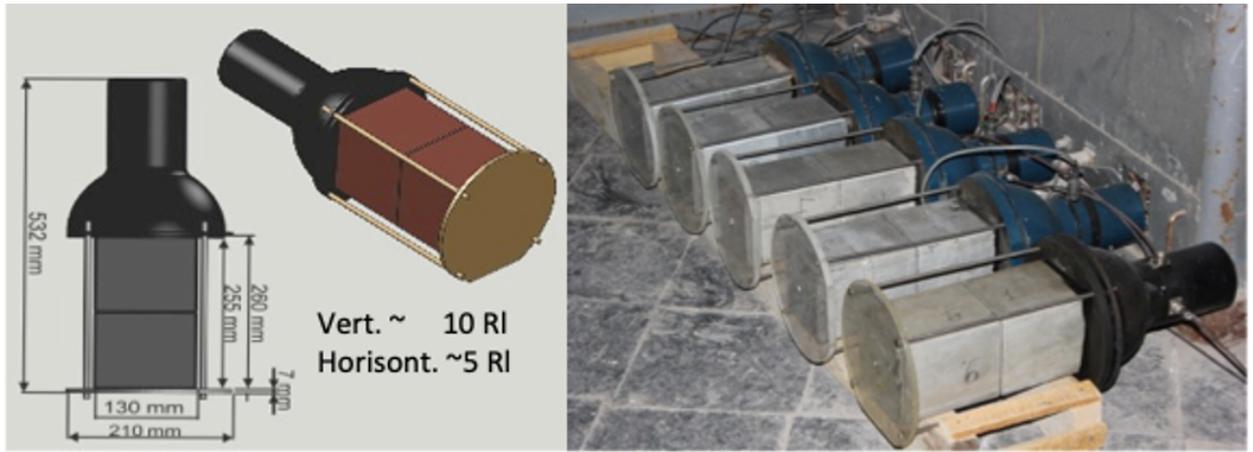

**Figure 2. NaI spectrometer layout and location under the roof of SKL hall under 0.7 mm of the iron tilt on Aragats station. The 130 mm of NAI (TL) scintillator corresponds to 5 radiation lengths (RL)**

The PMT signals are fed through a preamplifier to a logarithmic amplitude-to-digit-convertor (LADC). LADC specifications are described in detail in [8,9]. The scale factor **d** is a key to obtaining the digital code (energy deposit) of the particle:

$$K = d * \ln(E/E_o) + K_o \tag{1}$$

Where K is the code corresponding to PM amplitude, $K_o$ is the code obtained at ADC calibration using isotope emitting gamma rays with known energy $E_0$. To determine parameter **d**, we need at least two calibration points. For calibration, we use $^{137}$Cz (662 keV) and 60Co (1.12 MeV) isotopes (Fig. 3), and the so-called "muon peak", the peak in energy release distribution corresponding to CR muons ~60MeV (Fig. 4). In Fig.5 we show that the linearity interval covers energies from 0.3 to 60 MeV.

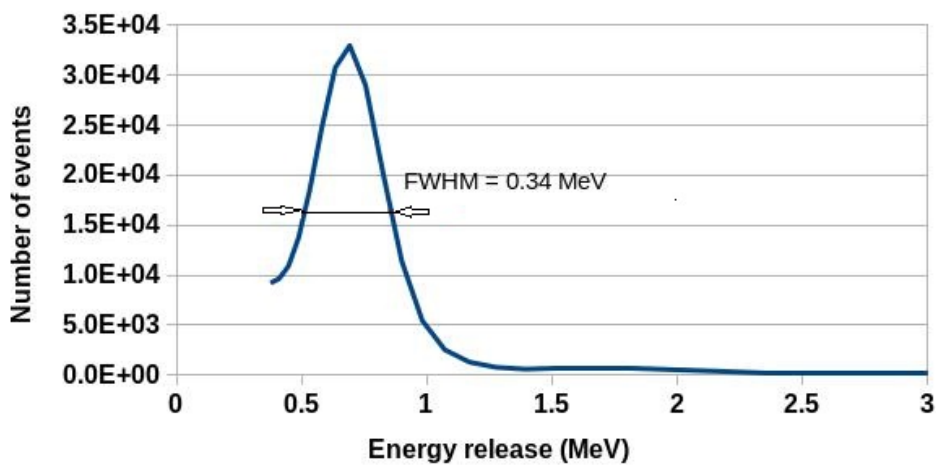

*Figure 3 Energy release distribution in the large home-made NaI spectrometer from exposing of $^{137}$Cz (662 кэВ), energy resolution for $^{137}$Cz isotope can be estimated as 340/662 ≈ 50%.*

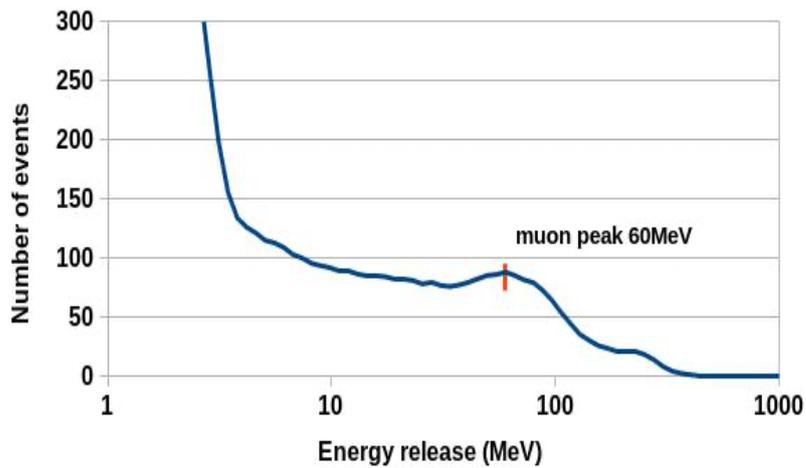

*Figure 4 Energy release distribution in NaI spectrometer from background cosmic rays.*

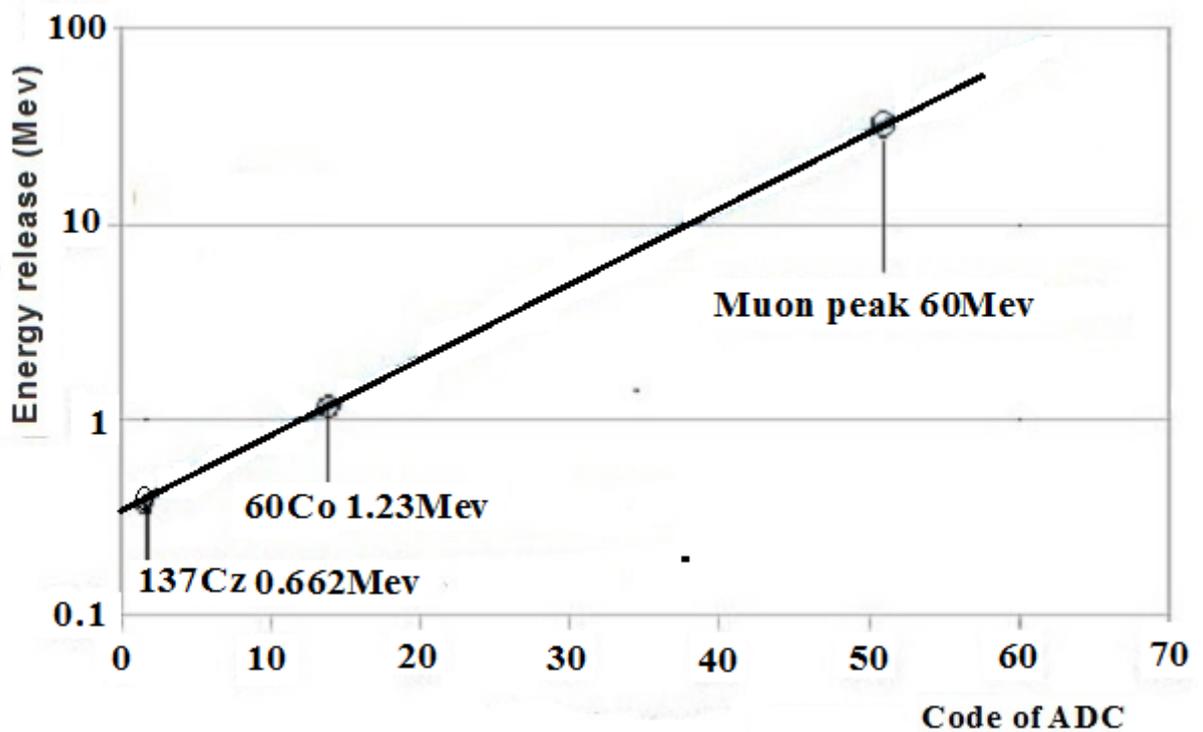

**Figure 5 Relation Energy release – ADC code for NaI spectrometer obtained from the calibration shown in Figs. 3 and 4.**

In Fig. 6 we show, that the resolution of ORTEC spectrometer is 7.3%, that allows to resolve as well gamma radiation of $^{60}$Co isotope with energies of 1.17 and 1.33 MeV.

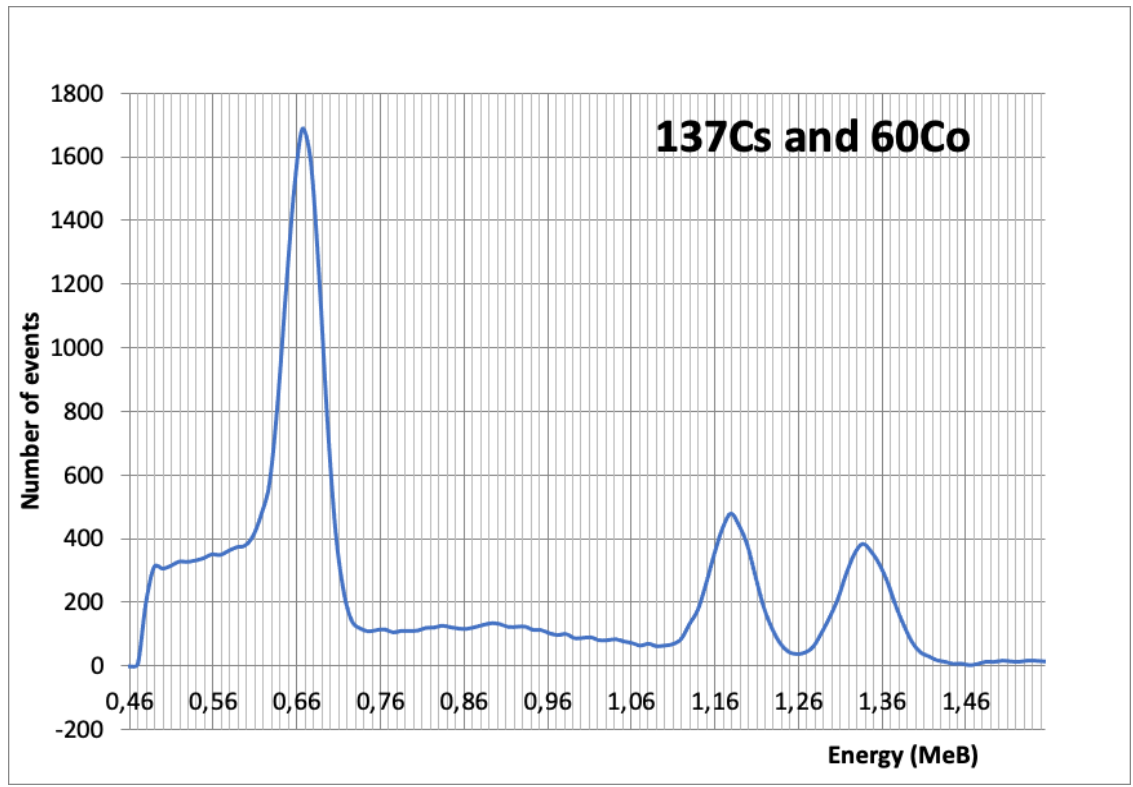

*Figure 6.* **Energy release distribution in NaI(TL) spectrometer produced by ORTEC company from exposing of $^{137}$Cz (662 кэB), and 60Co (1.17 and 1.33 MeV) isotopes, FWHM for the Cz isotope is ≈ 50 KeV, energy resolution for Cz isotope can be estimated as 48/662 ≈ 7.3%.**

3. **The data acquisition system for SEVAN**

The Advanced Data Acquisition System (ADAS, [10]) developed for particle detectors network is optimized for autonomous operation in remote places with no or minimal maintenance which also can be performed by technical staff without any IT skills. ADAS provides interfaces for the readout of experimental data from the NaI network, online monitoring, and control of detector parameters. All operations are executed remotely from institute headquarters either programmatically or via provided web interface.
ADAS software is distributed as preconfigured flashcards and is running on fanless mini PC computers (ADAS stations) without mechanical parts reducing the risk of hardware failures. The stations are fully operational with these cards and do not need any manual configuration. Each research station is provided with multiple spare flash cards and the technical personnel can simply replace the flash card in case of a software failure detected by specialists in Yerevan. Such failures are e.g. frequently caused by file system corruption due to frequent power outages. In winter, we also keep a spare MiniPC in the station to handle also hardware failures autonomously.
Software running at ADAS stations is continuously sampling the data from the NaI detectors and storing it as files in the local storage. These files are made available via HTTP protocol to the integration software running at the data center in Yerevan and possibly other locations. This software periodically queries all ADAS stations for new data and downloads it using rsync protocol. The downloaded raw files are archived. The information stored in the files is parsed and uploaded to the MySQL database for further distribution and data analysis via Advanced Data Extraction Infrastructure (ADEI) platform [11]. This architecture ensures high tolerance to network failures which are often the case due to severe weather conditions in the mountains. In the case of normal operation, the standard data latency between readout from detector and availability in public ADEI interface does not exceed 5 minutes.
To ensure data safety we rely on several levels of fault tolerance in
the data center. Several servers are querying data from ADAS stations
independently and maintaining individual archives of raw data and replicas of the MySQL database. Independent operation significantly simplifies design as no costly and complex synchronization between servers is required. All servers obtain the same data fully independently. One of the servers is designated as the master and serves the data to the clients.

In case of failure, a new system is elected and starts to serve data to
users transparently via Domain Name Service (DNS) mechanisms. The servers are located in 2 different locations in Armenia and additionally, a partial data mirror is maintained in Germany with our partners at Karlsruhe Institute of Technology (KIT). To further increase fault tolerance, all servers are using a mirroring RAID system to store both the raw data and database [12].

ADAS stores data in XML format along with metadata describing data
layout. The special conditions that occurred during the data acquisition also are stored in XML format. The layout
metadata helps to keep an uninterrupted track of the data channels while still allowing occasional reconfiguration of detectors including changes of parametrization affecting counting rates and also remapping of the detector channels. Operational metadata is intended to track operation failures that might affect detector operation and its readings. The metadata is stored in raw XML files, but it is not propagated to the MySQL database for performance reasons. Instead, it is used for automated data parsing and is also available for IT engineers for debugging purposes and for scientists to make more detailed investigations of particular events.

ADAS stations take care of the readout of experimental data from the
NaI network. Preliminary analysis is performed by providing operators with a minimalist web interface for monitoring and control. The web frontend delivers to the operator a full set of remote management capabilities including the possibility to monitor current data and adjust electronics. Along with displaying experimental measurements, ADAS provides a set of interfaces for controlling both detector electronics and software, including the remote tuning of the high voltage of photomultipliers (PM). The main ADAS page is shown in Fig. 7. From the left panel the real-time data and/or stored data can be chosen, i.e., count rates and energy releases histograms for further analysis and consequent electronics parameters tuning.

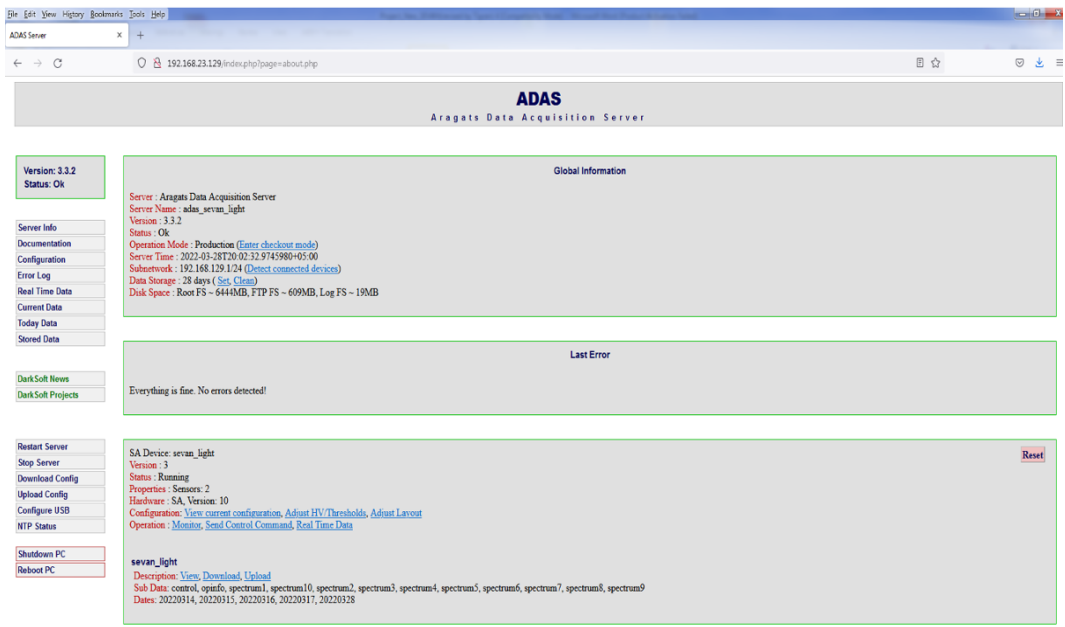

**Figure 7.** The ADAS software main menu.

4. **Energy spectra recovering**

During the multiyear monitoring of secondary cosmic rays on Aragats more than half-thousand thunderstorm ground enhancement (TGE) was registered by the NaI (Tl) and scintillation spectrometers. The TGE spectra were from distributions of ADC codes collected every 20 seconds for the large 4 m$^2$ area and 60 cm thick ASNT scintillation spectrometer ( see details of spectra recovery in [6]) or every one minute for of NaI (Tl) spectrometer and 0.25 m$^2$ and 20 cm thick CUBE scintillation spectrometer. As we can see in Fig. 6 a-d where differential energy spectra of 4 TGEs observed in 2018-2022 are depicted the recovering procedure is reliable and all 3 energy spectra well coincide.

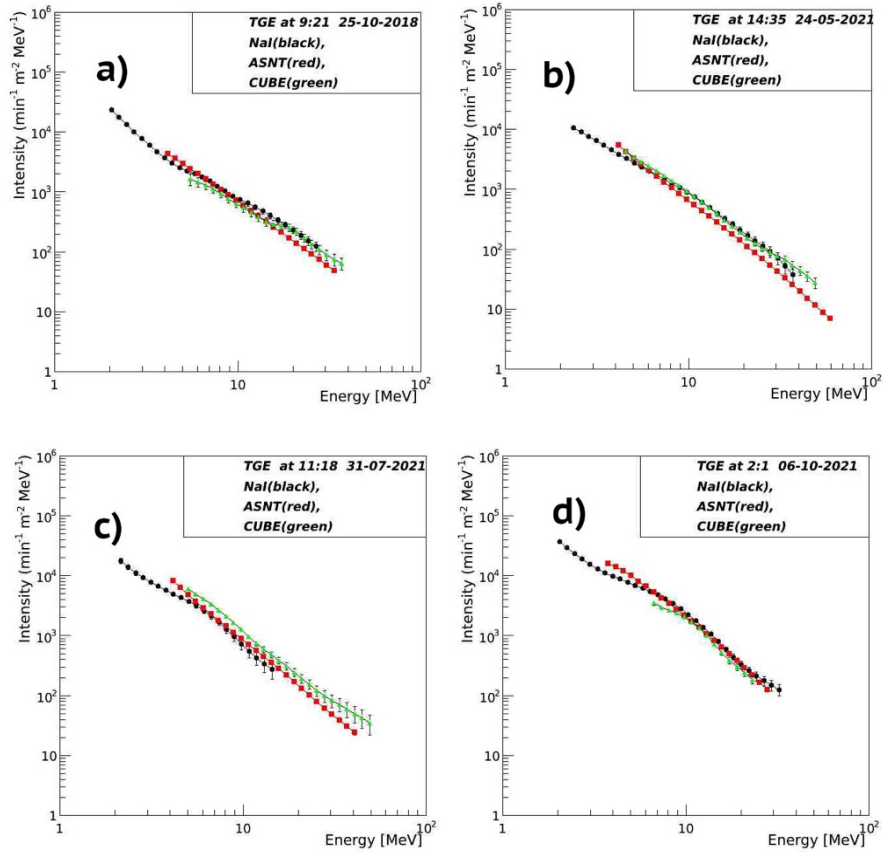

**Figure 8. Spectra of reconstructed TGEs from three ASEC detectors – NaI (black), ASNT (red) and CUBE (green).**

In Table 1 we show the statistics of power low fit parameters obtain with 4 energy spectra shown in Fig. 8. The relative error of both parameters do not exceed 20%, that is rather good coincidence, for the energy spectra of TGE particles from the RREA avalanches developing in the fast changing atmospheric electric field.

*Table 1. Parameters of power law fit dN/dE = A\*E $^{-\gamma}$ and their relative errors*

|  | Intensity A | Index γ | Intensity A | Index γ |
|---|---|---|---|---|
|  | 20 October 2018 | | 24 May 2021 | |
| ASNT | 1.02E+05 | 2.17 | 1.73E+05 | 2.36 |
| CUBE | 1.32E+05 | 2.12 | 1.47E+05 | 2.34 |
| NaI(Tl) | 1.23E+05 | 2.34 | 1.41E+05 | 2.21 |
| rel.error(%) | 12.94 | 5.22 | 11.07 | 3.54 |
|  | 31 July 2021 | | 6 October 2021 | |
| ASNT | 4.56E+05 | 2.59 | 1.83E+06 | 2.91 |
| CUBE | 3.05E+05 | 2.41 | 1.68E+06 | 2.91 |
| NaI(Tl) | 3.60E+05 | 2.71 | 1.26E+06 | 2.71 |
| rel.error(%) | 20.45 | 5.88 | 18.59 | 4.06 |

## 5. Horizontal profile of the atmospheric electric field

During a large storm on the First of May 2022, NSEF disturbances occurred both on Aragats and in Nor Amberd, reflecting the huge size of the storm, see Fig. 9a and 9b. In Fig.9b we show the thundercloud coverage of Armenia and, especially, its presence above the Aragats research station we estimate by the map of lightning locations with Boltek's Storm Tracker (lightning detection system [13]), powered by the software from Astrogenics. Storm tracker defines four types of lightning types: CG- (cloud-to-ground negative, lowering negative charge from the cloud to the ground), CG+ (cloud-to-ground positive, lowering positive charge to the ground), IC+ (normal polarity, intracloud, lowering positive charge to the ground) and IC -, (inverted polarity, intracloud, lowering negative charge from the cloud to the ground) in radii up to 480 km around the location of its antenna.

By the examining time-slices of the lightning activity we determine from which direction the storm is coming, and, finally, by putting them on the map of all lightning occurrences we can see if the storm's active zone misses the station (Fig. 9b).

Storm starts in Nor Amberd at 12:00 and on Aragats around a few minutes later (start of disturbances of the NSEF) and finished at ≈14:00. During the storm TGEs were observed both on Aragats and in Nor Amberd, see Fig. 10. On Aragats we use 60 cm thick and 4 m$^2$ area scintillation spectrometer ASNT[6], and in Nor Amberd ≈100 times smaller NaI spectrometer. Both are located under a 0.8 mm iron roof and register gamma rays and electrons with energies above 4 MeV. The significance of the peak observed in Nor Amberd was 9%, corresponding to 12σ. The significance of the first peak on Aragats at 13:01 was 3.7%, corresponding to 9σ, and, of the second peak, not seen in Nor Amberd, on 13:30 – 8%, corresponding to 20σ. Although there were no exact coincidences in TGE times, however, inside the huge thundercloud above Aragats mountain slopes, simultaneously, for a few tens of minutes electron accelerator sends high-energy avalanche particles in direction of the earth's surface.

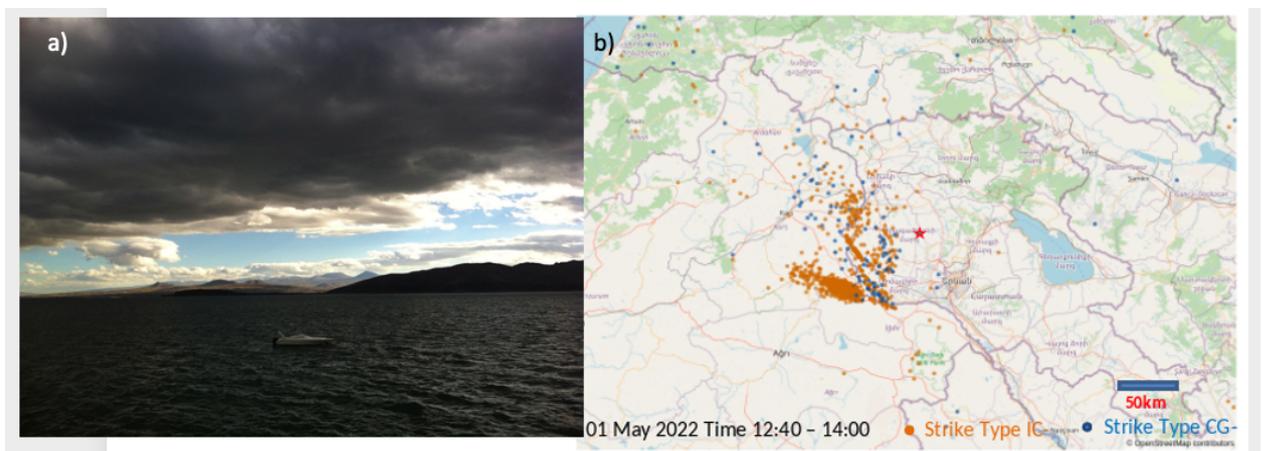

**Figure 9.** a) Thunderclouds above Aragats; b) mapping of lightning flashes registered during a storm coming from the south-west on May 1, 2022 at 12:40 – 14:00, coinciding with the time of enhanced electron and gamma fluxes. By a red star, the Aragats station location is denoted.

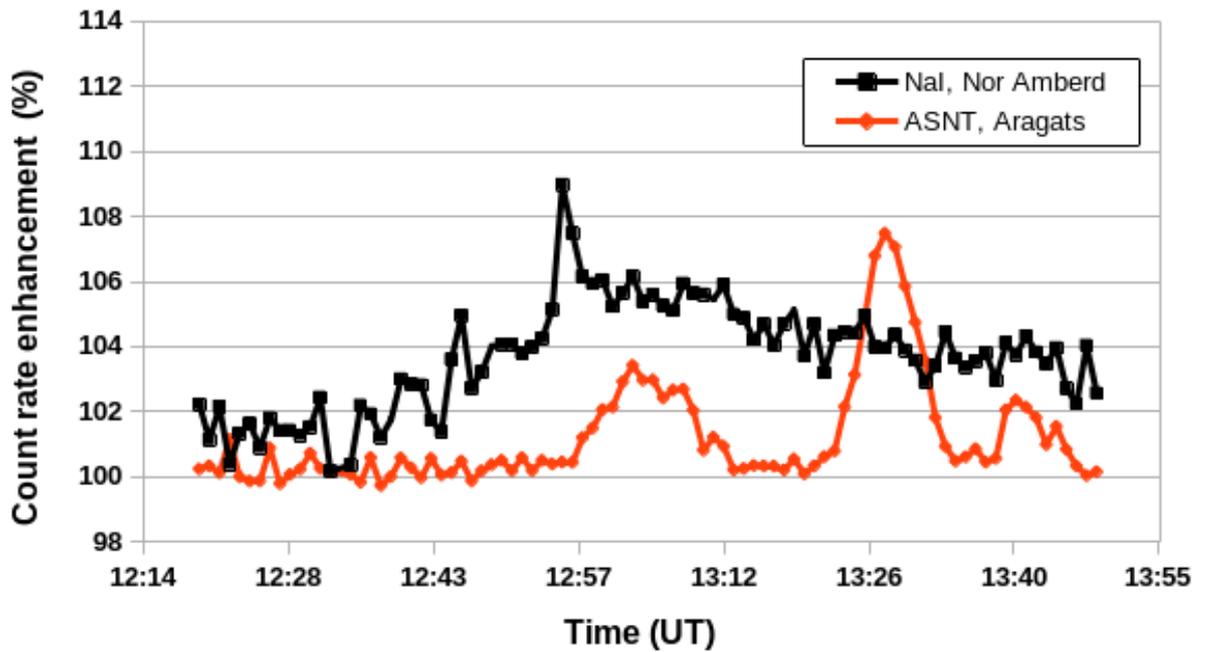

**Figure 10. 1-minute count rates of particle detectors on Aragats and in Nor Amberd (13 km apart**

In Fig.11 we show disturbances of NSEF, distances to the lightning flashes, and 1-minute time series of count rates of Aragats and Nor Amberd detectors. Large enhancements of the count rated occurred on Aragats at 13 – 13:14 and 13:23-13:33, in Nor Amberd at 12:30-13:23. The lightning active zone was far from Aragats, more than 15 km; a lot of flashes occurred nearby Nor Amberd, the nearest ones – at a distance of 1.5 km.

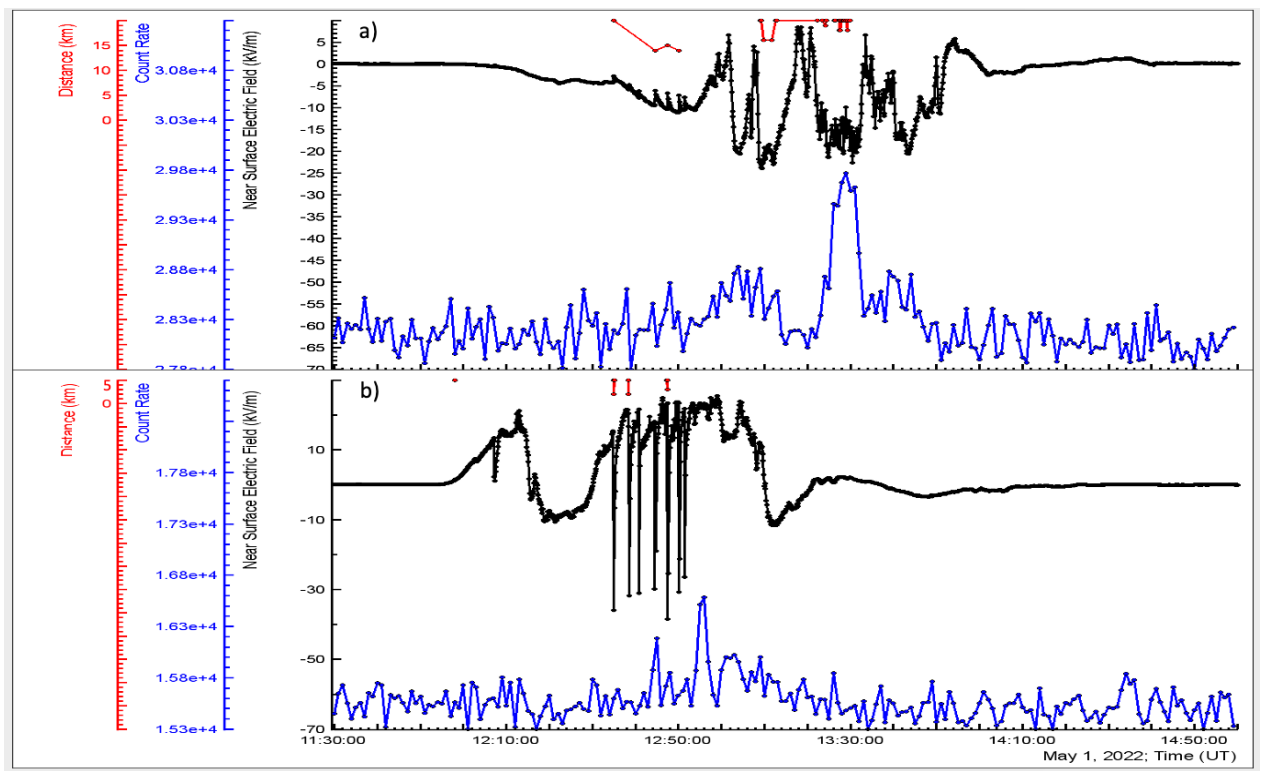

**Figure 11.** The disturbances of the NSEF (black); 1-minute count rates of 5 cm thick and 1 m$^2$ area plastic scintillators (blue); and distances to lightning flashes (red) measured on Aragats a) and in Nor Amberd b).

The NSEF changes during TGE from -23 to 8 kV/m on Aragats, and from -25 to 25 kV/m in Nor Amberd. The TGE occurred on Aragats during the deep negative electric field, and in Nor Amberd during positive NSEF. Thus, in spite of rather different conditions of the NSEF disturbances, and different charge structures in the cloud above, the TGEs in both destinations, share the same time and enhancement. Sure, we check the TGEs also by other particle detectors, on Aragats with solar neutron telescope, NaI spectrometer, 1 and 3 cm thick plastic scintillators, and in Nor Amberd with a tray of Geiger counters located outdoors.

All detectors measure approximately the same parameters of TGE. In Fig. 25 we show the time series of the maximum energies of the TGE measured each minute by the scintillation spectrometer ASNT on Aragats and – by a NaI(TL) spectrometer in Nor Amberd.

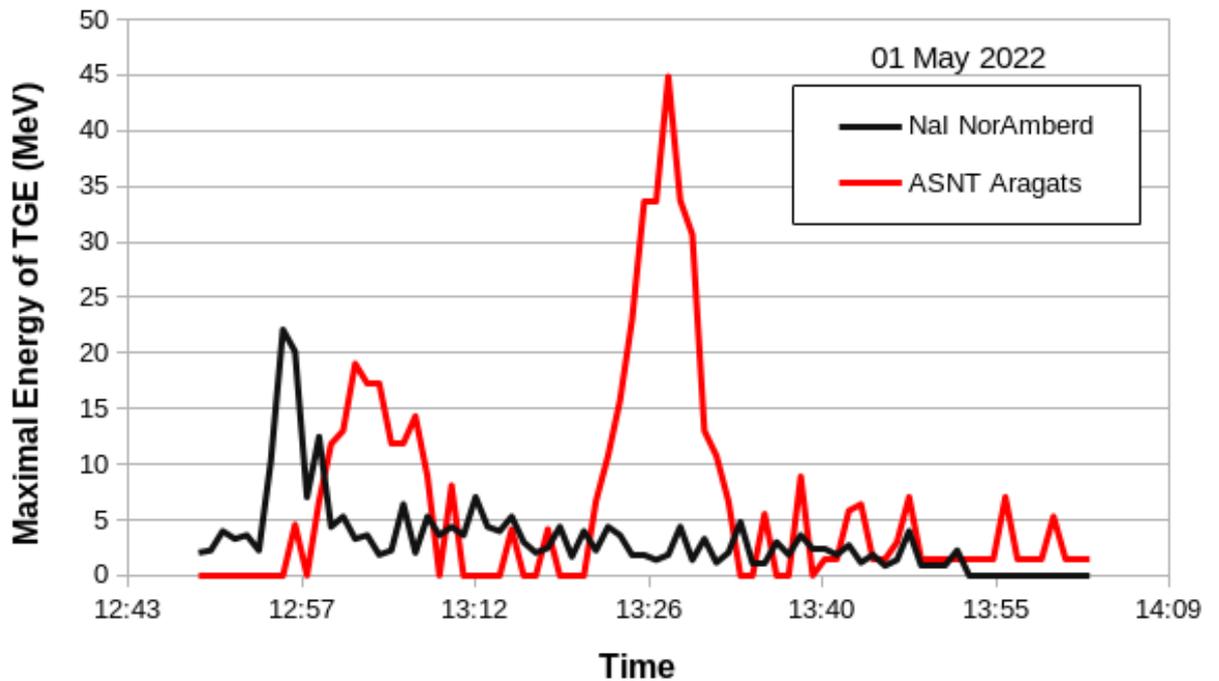

**Figure 12. The histogram of maximum energies of differential energy spectra measured each minute on Aragats with a large scintillation spectrometer ASNT (red), and in Nor Amberd by the NaI (TL) spectrometer.**

We can see coinciding in time peaks in the maximum energy distribution measured by both detectors. As we mention above NaI is a small detector with a size of only ≈0.03 m$^2$, and possibly it misses the detection of very rare high-energies, which ASNT reliably detects. And sure, conditions of TGE development in the skies above Aragats and Nor Amberd are different due to location: Aragats is on high land, and Nor Amberd is on the mountain slope; thus, the thundercloud can be very low above Aragats, and rather high above Nor Amberd. Nonetheless, as we demonstrate above, the particle fluxes were unleashed in a huge area of 12.8 km in distance and 1.5 km in height.

**Conclusions**

NaI spectrometers network operated on Aragats reliably recover differential energy spectra of TGE particles. Good agreement of the spectra recovered by the 3 spectrometers demonstrates that all 3 spectra coincide by 15%.
The horizontal profile of the atmospheric electric field on Aragats during thunderstorms can extend up to 10 km horizontally providing electron acceleration in the atmosphere, corresponding TGEs can cover at least 10 km$^2$ area on the earth's surface.

From the integral energy spectra of a TGE registered on May 30, 2018 [14], we estimate the total number of gamma rays (with energies above 300 keV) hitting the earth's surface to be 1.3*10$^6$/m$^2$min. Assuming that ≈ 2000 thunderstorms are active on the globe and that the overall surface of the thunderous atmosphere each moment can be estimated as 2.000 * 10 km$^2$ = 200,000 km$^2$ (0.04% of the globe surface), we come to an estimate of ≈10$^{15}$ gamma rays are hitting the earth's surface each second. Much more gamma rays of the same energy are accelerated by the upper dipole and bremsstrahlung gamma rays are directed to the open space.

## Acknowledgments

We thank the staff of the Aragats Space Environmental Center for the uninterruptable operation of all particle detectors and field meters, authors thank Danielyan Varugan for designing of the DAQ electronics. The authors acknowledge the support of the Science Committee of the Republic of Armenia (research project № 21AG-1C012) in the modernization of the technical infrastructure of high-altitude stations.

## Declaration of Competing Interest

The authors declare no conflict of interest.

## Data Availability Statement

The data for this study are available in numerical and graphical formats on the WEB page of the Cosmic Ray Division (CRD) of the Yerevan Physics Institute, http://adei.crd.yerphi.am/adei

## References

[1] Torii, T., Sugita, T., Kamogawa, M., Watanabe, Y., & Kusunoki, K. Migrating source of energetic radiation generated by thunderstorm activity. Geophysical Research Letters, 38 24 (2011).

[2] A.Chilingarian, G. Hovsepyan, G.Karapetyan, and M.Zazyan, Stopping muon effect and estimation of intracloud electric field, Astroparticle Physics 124 (2021) 102505.

[3] A Chilingarian, N Bostanjyan, T Karapetyan, On the possibility of location of radiation-emitting region in thundercloud, IOP Publishing Journal of Physics: Conference Series 409 (2013) 012217 doi:10.1088/1742-6596/409/1/012217

[4] Wada, Y., Enoto, T., Kubo, M., Nakazawa, K., Shinoda, T., Yonetoku, D., et al. (2021). Meteorological aspects of gamma-ray glows in winter thunderstorms. Geophysical Research Letters, 48, e2020GL091910. https://doi. org/10.1029/2020GL091910

[5] Hisadomi, S., Nakazawa, K., Wada, Y., Tsuji, Y., Enoto, T., Shinoda, T., et al. (2021). Multiple gamma-ray glows and a downward TGF observed from nearby thunderclouds. Journal of Geophysical Research: Atmospheres, 126, e2021JD034543. https://doi. org/10.1029/2021JD034543

[6] A. Chilingarian, G. Hovsepyan, T.Karapetyan, et al., Measurements of energy spectra of relativistic electrons and gamma-rays avalanches developed in the thunderous atmosphere with Aragats Solar Neutron Telescope, Journal of Instrumentation, 17 P03002 (2022).

[7] https://www.ortec-online.com/products/radiation-detectors/scintillation-detectors/scintillation-detector-types/905-series.

[8] G.Hovsepyan for ANI Collaboration  The measurement of the EAS charged particle component  using a logarithmic ADC.  ANI98,  FZKA 6215, Forschungszentrum Karlsruhe, 1998, p.45

[9] K. Arakelyan, A.Daryan,  G.Hovsepyan, L. Kozliner,  A.Reimers    Design and response function of NaI detectors of Aragats complex installation  Nuclear Instruments and Methods in Physics Research  A763 (2014) 308–313.

[10] S. Chilingaryan, "Universal Data Exchange Solution for Modern Distributed Data Acquisition Systems and Its Implementation for Cosmic Ray Monitor Networks," in Institute for

Informatics and Automation Problems of National Academy of Science of the Republic of Armenia, Yerevan, vol. Ph.D. Thesis, 2006.

[11] S. Chilingaryan, A. Beglarian, A. Kopmann, S. V̈ocking, Advanced Data Extraction Infrastructure: Web Based System for Management of Time Series Data, Journal of Physics: Conference Series **219,** (2010), 042034, doi:10.1088/1742-6596/219/4/042034.

[12] S.Chilingaryan, A.Chilingarian, V.Danielyan, W.Eppler, Advanced data acquisition system for SEVAN, Advances in Space Research, 43, (2009), 717–720, doi:10.1016/j.asr.2008.10.008.

[13] https://www.boltek.com/product/ld350-long-range-detection-kit

[14] A.Chilingarian, G. Hovsepyan, S. Soghomonyan, M. Zazyan, and M. Zelenyy, Structures of the intracloud electric field supporting origin of long-lasting thunderstorm ground enhancements, Physical review 98, 082001(2018).